\documentclass[aps,prl,nofootinbib,twocolumn,floatfix,letterpaper,superscriptaddress,showpacs]{revtex4}
\usepackage{graphicx}\usepackage{natbib}\usepackage{amsmath}\usepackage{color}
\pdfoutput=1
\usepackage{times}

\newcommand{\es}[2] {\begin{equation} \label{#1} \begin{split} #2 \end{split} \end{equation}}

\begin{document}

\title{Multi-PeV Signals from a New Astrophysical Neutrino Flux Beyond the Glashow Resonance}

\author{Matthew D. Kistler}
\email{kistler@stanford.edu}
\affiliation{Kavli Institute for Particle Astrophysics and Cosmology, Department of Physics, Stanford University, Stanford, California 94035 and SLAC National Accelerator Laboratory, Menlo Park, California 94025}

\author{Ranjan Laha}
\email{ranjalah@uni-mainz.de}
\affiliation{PRISMA Cluster of Excellence and
             Mainz Institute for Theoretical Physics,
             Johannes Gutenberg-Universit\"{a}t Mainz, 55099 Mainz, Germany}
\affiliation{Kavli Institute for Particle Astrophysics and Cosmology, Department of Physics, Stanford University, Stanford, California 94035 and SLAC National Accelerator Laboratory, Menlo Park, California 94025}

\date{\today}

\begin{abstract}
The IceCube neutrino discovery was punctuated by three showers with $E_\nu \!\approx\! 1\!-\!2$~PeV.
Interest is intense in possible fluxes at higher energies, though a deficit of $E_\nu \!\approx\! 6$~PeV Glashow resonance events implies a spectrum that is soft and/or cutoff below $\sim\,$few PeV.
However, IceCube recently reported a through-going track depositing $2.6 \!\pm\! 0.3$~PeV.
A muon depositing so much energy can imply $E_{\nu_\mu} \!\gtrsim\!10$~PeV.
Alternatively, we find a tau can deposit this much energy, requiring $E_{\nu_\tau}$$\sim\!10 \times$ higher.
We show that extending soft spectral fits from TeV--PeV data is unlikely to yield such an event, while an $\sim$$E_\nu^{-2}$ flux predicts excessive Glashow events.
These instead hint at a new flux, with the 
hierarchy of $\nu_\mu$ and $\nu_\tau$ energies implying astrophysical neutrinos at $E_\nu \!\sim\! 100$~PeV if a tau.  We address implications for ultrahigh-energy cosmic-ray (UHECR) and neutrino origins.
\end{abstract}

\maketitle

{\bf Introduction.}
The discovery of astrophysical neutrinos by IceCube  \cite{Aartsen2013,Aartsen2013b,Aartsen2014,Aartsen2015,Kopper2015,Schoenen2015,Aartsen2015b,Aartsen:2016ngq,Aartsen:2016xlq} allows for new characterizations of the high-energy universe.
Neutrinos can arise from cosmic-ray interactions within sources (e.g., \cite{Greisen:1960wc,Bahcall1964,Muecke:2002bi}) and with extragalactic photon backgrounds (e.g., \cite{Beresinsky:1969qj,Stecker:1978ah,Hill:1983mk,Yoshida:1993pt,Engel:2001hd,Yuksel2007,Gelmini:2011kg,Aloisio:2015ega}).  The fluxes vary greatly depending on assumptions and data may yield insight into the inner workings of UHECR accelerators \cite{Hillas:1985is} or unexpected physical effects  \cite{Pakvasa:2012db,Kistler2015d}.

Along with dozens of $\sim$10--100~TeV events, IceCube detected three contained-vertex showers with deposited energy $E_{\rm dep}$$\approx\,$1--2~PeV (likely with $E_{\nu}$$\approx\,$$E_{\rm dep}$)  \cite{Aartsen2013,Aartsen2014}.
The neutrino spectrum indicated below PeV energies is significantly softer than $E_\nu^{-2}$, reaching a sharp upper limit at $E_\nu\! \gtrsim\! 5$~PeV  (5$\times$$10^6$~GeV; Fig.~\ref{casca}) due to a lack of $\sim$6~PeV showers from on-shell $\bar{\nu}_e$$e$$\rightarrow$$W^-$ \textit{Glashow resonance}\,\cite{Glashow:1960zz} scattering.

However, IceCube recently reported an upgoing through-going track depositing $E_{\rm dep}$$=\! 2.6$$\pm$$0.3$~PeV \cite{Schoenen2015,Aartsen:2016ngq,Aartsen:2016xlq}.  We will see that the required $E_{\nu}$ to produce this event is $\gg$$E_{\rm dep}$, significantly larger than even the PeV shower events.
This highest-energy event raises important questions concerning astrophysical neutrinos, including, subtly: what flavor of neutrino produces such a track?

\begin{figure}[b!]
\includegraphics[width=1.01 \columnwidth,clip=true]{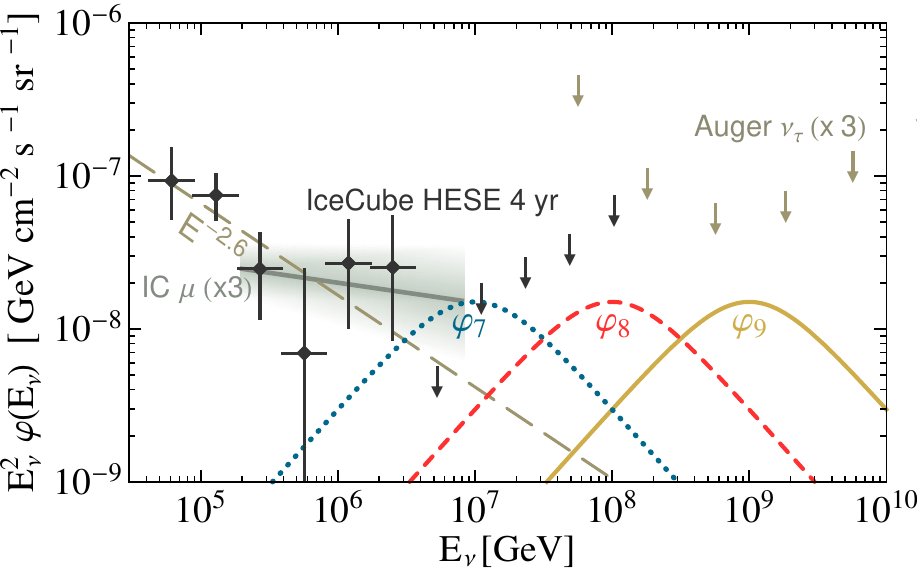}
\vspace*{-0.6cm}
\caption{
IceCube 4~yr contained HESE data \cite{Kopper2015} (which do {\it not} include the $E_{\rm dep} \!=\! 2.6$~PeV track event), IceCube 6~yr $\nu_\mu$ band (assumes the PeV track is a muon \cite{Aartsen:2016xlq}), and Auger $\nu_\tau$ upper limits \cite{Aab:2015kma}.
Also, an $E_\nu^{-2.6}$ flux ({\it long-dashed}) and extragalactic spectral models peaking near $10^7$~GeV ($\varphi_7$; {\it dotted}), $10^8$~GeV ($\varphi_8$; {\it dashed}), and $10^9$~GeV ($\varphi_9$; {\it solid}).  Models $\varphi_7$ and $\varphi_8$ resemble BL~Lac AGN models, while rescaled combinations of $\varphi_7$ and $\varphi_9$ approximate GZK neutrinos from EBL and CMB interactions.
All data and fluxes are summed over flavors (and $\nu$+$\bar{\nu}$), assuming $\varphi_{\nu_e} \!=\! \varphi_{\nu_\mu} \!=\! \varphi_{\nu_\tau}$ and $\varphi_{\nu} \!=\! \varphi_{\bar{\nu}}$.
\label{casca}}
\end{figure}

We first consider the standard assumption that the track is a muon.  We show: ($i$) soft astrophysical neutrino spectra (e.g., $E_\nu^{-2.6}$) are unlikely to produce such muons; ($ii$) harder spectra (e.g., $\sim$$E_\nu^{-2}$) overproduce Glashow shower rates.
This motivates us to better characterize the super-Glashow energy regime.  We examine heuristic spectral models covering a variety of production scenarios and their expected signals.

We also consider an intriguing possibility of a track left by a tau lepton.
Though detection methods for $\nu_\tau$ have been discussed over many years (e.g., \cite{Learned:1980jh,Learned:1994wg,Fargion:1997eg,Halzen:1998be,Dutta:2000jv,Beacom:2001xn,Beacom:2003nh,Jones:2003zy,Yoshida:2003js,Bugaev:2003sw,Dutta:2005yt,DeYoung:2006fg}), no distinct $\tau$-like event has yet been identified by IceCube~\cite{Aartsen:2015dlt}.  Energy deposition by taus within the detector leads to many possible signals (see \cite{DeYoung:2006fg}).  However, through-going tau tracks are little discussed and energy-loss stochasticity presents difficulty in individually identifying PeV tracks as muons or very-long-lived taus with decay length
$\gamma_\tau \, c \, \tau_\tau \!\approx\! (E_\tau/20$~PeV)~km.

For either scenario, we deduce a harder, higher-energy astrophysical neutrino flux than previously measured is more likely present.
A tau track traversing the $\sim$1~km detector without decaying would imply a much-higher parent neutrino energy, and give an unexpected window into astrophysical neutrinos at $\sim$100~PeV.
We address differences in the energy spectrum and angular distribution of tau and muon events and discuss implications for outstanding problems in UHECR and neutrino physics.

\begin{figure*}[t!]
\hspace*{-0.4cm}
\includegraphics[width=1.0 \columnwidth,clip=true]{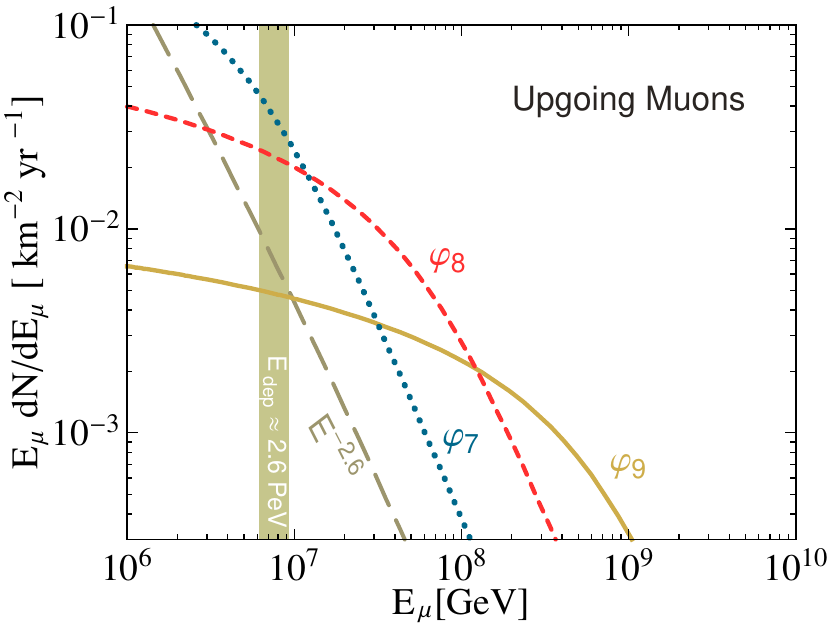}
\includegraphics[width=1.0 \columnwidth,clip=true]{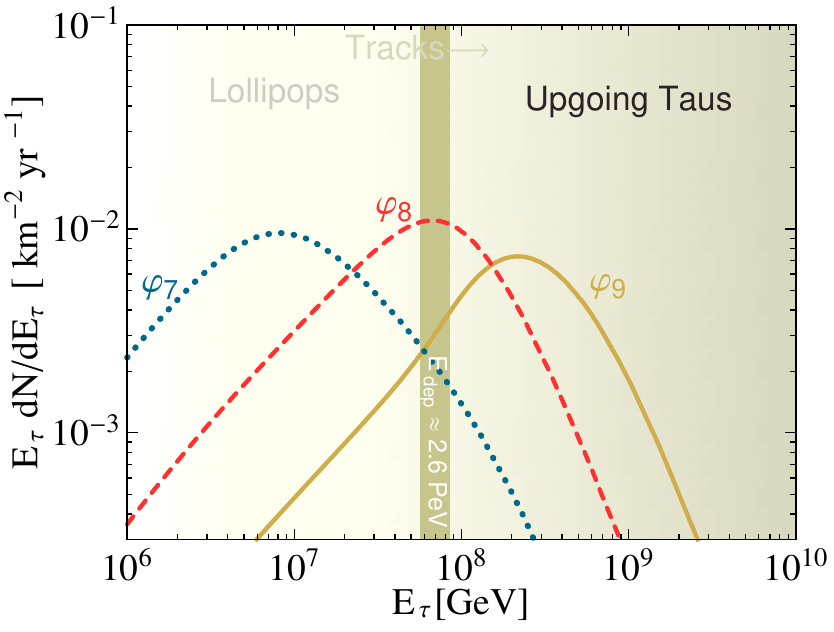}
%
\caption{
{\it Left:} Spectra of upgoing muons (with $E_\mu$ entering detector) from neutrino models in Fig.~\ref{casca}.  To deposit $\sim$2.6~PeV suggests $E_\mu \!\gtrsim\!8$~PeV (vertical band), with a $\gtrsim$10 PeV energy of the $\nu_\mu$.
{\it Right:} The same for taus, denoting ranges of dominant entering-tau event topologies.
Through-going tau deposition of $\sim$2.6~PeV suggests $E_\tau \!\gtrsim\!70$~PeV (vertical band), a much larger $E_\nu$ than a muon depositing the same energy.
}
\label{muonspec}
\end{figure*}

{\bf Multi-PeV Tracks.}
Analytic methods have been presented for charged- (CC) and neutral-current (NC) shower event rates in IceCube \cite{Kistler2014,Laha2013} and muon fluxes from $\nu_\mu$ interactions \cite{Gaisser:1990vg,Kistler2006,Beacom:2007yu}, though these cannot be directly applied to long-lived taus.

We determine the tau flux spectrum $dN_\tau/d E_\tau$ in ice using a volumetric source term $Q(E_\tau)$ for taus produced by $\nu_\tau$
\begin{equation}
  \frac{d}{d E_\tau} \left[ b_\tau(E_\tau) \frac{dN_\tau}{d E_\tau} \right] + \frac{m_\tau}{c\,\tau_\tau E_\tau} \frac{dN_\tau}{d E_\tau} = Q(E_\tau) \,,
\label{conteq}
\end{equation}
with tau energy loss $b_\tau(E_\tau) \!=\! dE_\tau/dX$, mass $m_\tau$, and lifetime $\tau_\tau$.
We find $b_\tau(E_\tau) \!=\! b_0\, \rho\, (E_\tau/{\rm GeV})^{\kappa_\tau}$, within density $\rho$ with $b_0 \!=\! -4.6 \times 10^{-9}\,$GeV~cm$^{2}$~g$^{-1}$ and $\kappa_\tau$$=\! 5/4$, adequately approximates parametrized Monte Carlo results of \cite{Dutta:2005yt} in our $E_\tau$ range of interest.  This form is simple to implement in solving Eq.~(\ref{conteq}) via an integrating factor solution (e.g., \cite{Arfken}).  After simplification, we obtain
\begin{eqnarray}
  \frac{dN_\tau}{d E_\tau} &=& \frac{1}{-b_\tau(E_\tau)} \exp\left[\frac{m_\tau}{c\, \tau_\tau \kappa_\tau\, b_\tau(E_\tau)}\right] \nonumber \\
                             & & \times \int_{E_\tau}^{E^{\rm max}} \!dE\, Q(E)\, \exp\left[-\frac{m_\tau}{c\, \tau_\tau \kappa_\tau\,b_\tau(E)}\right] 
     .
  \label{taueq}
\end{eqnarray}
For muons, the exponential terms vanish ($\tau_\mu \gg \tau_\tau$) and $b_\mu(E_\mu)$$=$$ -\alpha_\mu \!-\! \beta_\mu E_\mu$, using a stochastic loss fit \cite{Koehne:2013gpa}: $\alpha_\mu$$=\!2.49 \times 10^{-3}\,$GeV~cm$^2$~g$^{-1}$ and $\beta_\mu$$=\!4.22 \times 10^{-6}\,$cm$^2$~g$^{-1}$.

\begin{figure*}[t!]
%
\hspace*{-0.4cm}
\includegraphics[width=0.99 \columnwidth,clip=true]{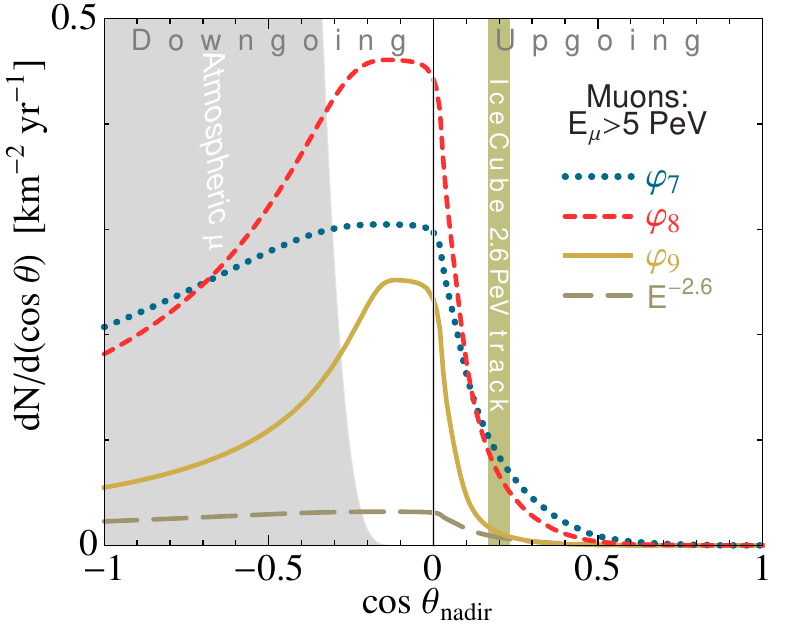}
\includegraphics[width=0.99 \columnwidth,clip=true]{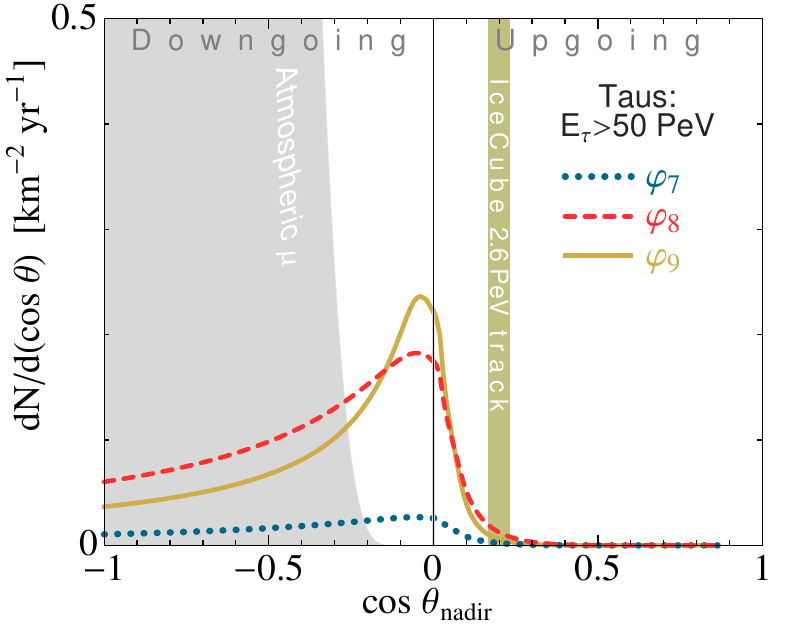}
%
%
\caption{
{\it Left:} Angular distribution of $E_\mu \!>\! 5$~PeV muons for neutrino models in Fig.~\ref{casca}.
{\it Right:} The same for $E_\tau \!>\! 50$~PeV taus.
The cutoffs towards larger upgoing angles are due to Earth attenuation, while the decline to larger downgoing angles is due to the finite ice depth.
Both are compared to the direction of the track event ($\theta_{\rm nadir} \!\approx\! 78.5^\circ$) and background atmospheric muons with $E_\mu \!>\! 5$~PeV at the detector ({\it shaded}).
\label{angspec}}
\end{figure*}

We first consider downgoing events, where fluxes are simpler.
At PeV and greater energies the differential $\nu N$ charged-current cross section $d\sigma_{\rm CC}/dy$ is strongly peaked at $y \!=\! 0$ \cite{Gandhi:1995tf}.  We use $E_\tau$=$ \langle 1-y \rangle E_\nu$, approximating $\langle 1-y \rangle$$=\! 0.8  \!=\! q$ (ignoring weak $E_\nu$ dependence  \cite{Gandhi:1995tf}),
\begin{equation}
  Q(E_\tau) \approx N_A\,\rho\, \varphi_\tau(E_\tau/q)\, \sigma_{\rm CC}(E_\tau/q)/q ,
\label{Qeq}
\end{equation}
where $N_A\,\rho$ is the molar density of ice.  We find this adequately approximates the birth spectrum of taus (and muons) using the differential cross section.

$E^{\rm max}$ relates the energy at the detector to a birth energy at the surface.
The particle range from arbitrary energy losses can be inverted (see \cite{Kistler2015}), though the $b(E)$ above allow for analytic solutions.
For taus, 
$E_\tau^{\rm max}$$=\! [E_\tau^{-1/4} + b_0 \ell(\theta)/4]^{-4}$,
 where $\ell(\theta)$ is the column depth to the surface at $\theta$ in cm~water-equivalent (we assume a 2~km depth).
For muons, 
$E_\mu^{\rm max}$$=\! \left\{{\rm exp}[ \beta_\mu \ell(\theta)] (\alpha_\mu \!+\! \beta_\mu E_\mu)-\alpha_\mu\right\}/\beta_\mu$.

For upgoing fluxes, effectively $E^{\rm max}$$\rightarrow$$ \infty$.
We use $\ell_\oplus(\theta)$ \cite{Dziewonski:1981xy} for attenuation, $e^{-\tau_\oplus}$, with $\tau_\oplus \!=\! N_A\, \ell_\oplus(\theta)\,\sigma_{\rm tot}(E_\nu)$.
For $\nu_e$ and $\nu_\mu$, $\sigma_{\rm tot} \!=\! \sigma_{\nu N}$, with $\sigma_{\rm tot} \!=\! \sigma_{\bar{\nu} N}$ for $\bar{\nu}_\mu$.  For $\bar{\nu}_e$ we must add $ \sigma_{\bar{\nu}_e e}$, which practically excludes a $W^-$$\rightarrow$$\mu^-\bar{\nu}_\mu$ origin of the 2.6~PeV track.

Upgoing $\nu_\tau$ fluxes are complicated by regeneration, decays of taus produced within Earth back into $\nu_\tau$.  The total $\nu_\tau$ number flux is conserved, although the spectrum is distorted towards lower $E_{\nu_\tau}$.
We estimate the surviving $\nu_\tau$ flux by converting the interacting fraction for each $E_{\nu_\tau}$ into a continuous distribution based on \cite{Bugaev:2003sw} (neglecting regenerated $\nu_\mu$/$\nu_e$).

{\bf  Super-Glashow Fluxes.}
$E_{\nu}$ probed by a fully-through-going track event depends on the parent neutrino flavor.
If the 2.6~PeV track event is from a muon, estimating $E_{\rm dep}$ in $\sim$1~km by integrating $b_\mu(E_\mu)$ implies $E_\mu \!\gtrsim\! 8$~PeV upon entering IceCube (Fig.~\ref{muonspec}; {\it left}).

Compared to a muon with the same energy, the energy loss rate of a tau is much smaller.  Depositing $E_{\rm dep} \!=\! 2.6$~PeV in $\sim$1~km from $b_\tau(E_\tau)$ alone (i.e., not including any energy from the $\nu_\tau$ interaction or tau decay, both assumed to occur outside the detector) implies $E_\tau \!\approx\! 67$~PeV.  The light yield may even be less than a muon of this $E_{\rm dep}$ dependent upon photonuclear losses \cite{DeYoung:2006fg}.
Since $E_\tau \!\gg\! E_\mu$, the difference in neutrino energy required for a through-going tau track is significant.

Fig.~\ref{muonspec} shows spectra of muons ({\it left}) and taus ({\it right}) versus energy entering the detector.
We see that an $E_\nu^{-2.6}$ spectrum similar to IceCube fits \cite{Aartsen2015,Kopper2015} (Fig.~\ref{casca}) implies a very-low rate of multi-PeV muons (and a negligible tau rate not shown).
A prompt PeV neutrino flux should be steeper with a lower normalization than the $E_\nu^{-2.6}$ model  \cite{Kopper2015,Bhattacharya:2015jpa,Laha:2016dri}, with $<\!0.01$\% probability of an atmospheric origin for the track event \cite{Schoenen2015,Aartsen:2016ngq,Aartsen:2016xlq}.
A quantitative comparison with plausible astrophysical models can provide flux levels yielding more adequate rates.

The neutrino spectrum from $pp$ scattering roughly traces the proton spectrum within the source.  Spectra from $p\gamma$ scattering, set by protons and target photons above the photopion threshold, tend to be hard prior to being broken and/or cutoff.

We consider spectra to examine super-Glashow neutrino flux levels at Earth described as
\begin{equation}
      \varphi_i(E_\nu) =   f_i  \left[\left(E_\nu/E_i\right)^{\alpha \eta} + \left(E_\nu/E_i\right)^{\beta \eta} \right]^{1/\eta} \,,
\label{specfit}
\end{equation}
with $\alpha \!=\! -1$, $\beta \!=\! -3$, broken at $E_i \!=\! 10^7$, $10^8$, and $10^9$~GeV corresponding to Models~$\varphi_7$, $\varphi_8$, and $\varphi_9$, respectively, with $\eta \!=\! -1$ to smoothly mimic source variation and cosmic evolution.
One could instead use exponential cutoffs, though the spectral peak, rather than high-energy tail, mostly sets rates.

The $\varphi_i$ spectra (Fig.~\ref{casca}) use equal peak normalization, though each can be rescaled and/or summed for model-dependent descriptions (e.g., \cite{Stecker:2013fxa,Padovani:2015mba,Kistler2015c,Baerwald:2013pu,Murase2013}).
Model $\varphi_7$ peaks near $E_{\nu_\mu}$ for a minimal muon interpretation of the 2.6~PeV track.
It also approximates the $p \gamma$ spectral shape in High-energy-peaked BL Lac (HBL) AGN models, while $\varphi_8$ resembles Low-energy-peaked BL Lac (LBL) \cite{Muecke:2002bi,Padovani:2015mba}.
Model $\varphi_9$ approximates the GZK (cosmogenic) neutrino spectrum from $p \gamma$ interactions on the CMB and $\varphi_7$ for lower-energy proton interactions with the extragalactic background light (EBL), which can be combined for various cosmogenic scenarios \cite{SuppM}.

{\bf Multi-PeV Rates.}
Fig.~\ref{muonspec} shows upgoing muon and tau spectra from $\varphi_i$ models (Fig.~\ref{casca}).
Muon and tau energy deposition are more or less stochastic (e.g., \cite{Koehne:2013gpa,Aartsen:2013vja}).
For concreteness, we consider $E_\mu \!>\! 5$~PeV and $E_\tau \!>\! 50$~PeV rates (and in Fig.~\ref{angspec}).
This still corresponds to tau energies allowing traversal of IceCube before decaying.   

Downgoing muons and taus are also relevant from the angular region where background is low enough to safely assume an astrophysical origin.
A PeV muon flux is expected from atmospheric cosmic-ray interactions.
We estimate this background relating the muon spectrum at the surface to that reaching the detector accounting for energy loss (e.g., \cite{Gaisser:1990vg}).  Being concerned with PeV energies and above, we use a spectrum approximating prompt muons \cite{Aartsen:2015nss}, $dN/dE_\mu \propto E_\mu^{-3}$, neglecting muon bundles (discussed by IceCube \cite{Aartsen:2015nss}).
Fig.~\ref{angspec} shows the angular distribution of atmospheric muons with $E_\mu \!>\! 5$~PeV at detector depth.
The ice effectively eliminates these $\lesssim\! 10^\circ$ above the ``horizon''.

Fig.~\ref{angspec} compares the angular distributions of $E_\mu \!>\! 5$~PeV muons and $E_\tau \!>\! 50$~PeV taus.
Table~\ref{tab:rates} shows rates in 5~km$^2\,$yr, with showers for 5~km$^3\,$yr calculated as in \cite{Kistler2014,Laha2013},
including downgoing tracks within $-0.2 \!<\,$cos$\,\theta_{\rm nadir} \!<\! 0$.
Adding to upgoing rates yields $\sim$0.5--1 one total muon/tau track for each of $\varphi_7$, $\varphi_8$, and $\varphi_9$, while $E_\nu^{-2.6}$ remains small.
We see for $\varphi_7$$\rightarrow$$\varphi_8$$\rightarrow$$\varphi_9$ the tau/muon track ratio approaches unity.

The Fig.~\ref{muonspec} spectra do not attempt to correct for IceCube energy resolution.  While for muons this is fairly straightforward, with reconstruction yielding better resolution at high energies \cite{Aartsen:2013vja}, for taus the correspondence between energy and decay length complicates event topologies.  Fig.~\ref{muonspec} illustrates energies characteristic of entering-tau classes: ``lollipops'' in which a tau enters the detector and decays (i.e., in its last $\sim$1~km), transitioning (via shading) to ``tracks'' traversing the entire detector.  Overestimating $E_\tau$, for instance, does not result in an increase in actual range and would not change the topology.

The energies required to deposit $\sim\! 2.6$~PeV calculated here are indicative.
Uncertainty in tau photonuclear losses affects the visible signal \cite{DeYoung:2006fg} and a more thorough investigation should be carried out by IceCube.
Even with a more precise calculation, our conclusion will remain valid: the energy of a tau must be much larger than that of a muon in order to deposit the same amount of track energy.  The $\tau$-track signal is often neglected (cf., \cite{Learned:1980jh}), and even if this track turns out to favor a muon, we encourage optimizing tools for through-going taus.

\begin{table}[t!]
\caption{Events in 5~km$^2\,$yr (tracks: $E_{\mu}$$>$$5$~PeV or $E_{\tau}$$>$$50$~PeV; upgoing or downgoing within cos$\,\theta_{\rm nadir}$$>$$-0.2$) and 5~km$^3\,$yr (showers: $E_{{\rm em}}$$>$$5$~PeV).}
\label{tab:rates}
\begin{ruledtabular}
\begin{tabular}{lccccccc}
					  & $E_\nu^{-2.13}$	& $E_\nu^{-2.13}$e & $E_\nu^{-2.6}$ & $E_\nu^{-2.6}$c 	& $\varphi_7$	& $\varphi_8$	& $\varphi_9$ \\ \hline
upgoing $\mu$	 		  & 0.05		       	& 0.04			   & 0.05			& 0.02			& 0.22		& 0.25		& 0.08 \\
down $\mu$			  & 0.05			& 0.04			   & 0.08			& 0.01			& 0.30		& 0.46		& 0.25 \\
upgoing $\tau$			  & ---			& ---				   & ---			& ---				& 0.01		& 0.08 		& 0.07 \\
down $\tau$			  & ---			& ---				   & ---			& ---				& 0.03		& 0.17		& 0.19 \\ \hline
track sum				  & 0.1		 	& 0.08			   & 0.13			& 0.03			& 0.56		& 0.96		& 0.59 \\ \hline
$\bar{\nu}_e e$ shower	  & 3.0			& 1.6				   & 1.0 			& 1.0				& 2.6 		& 0.36		& 0.04 \\
$\nu_e\!+\!\bar{\nu}_e$ CC & 0.48			& 0.28			   & 0.26			& 0.16			& 0.87		& 0.50		& 0.12  \\
$\nu+\bar{\nu}\,$  NC	  & 0.01			& 0.01			   & 0.05			& 0.0				& 0.18		& 0.42		& 0.16 \\
\end{tabular}
\end{ruledtabular}
\end{table}

{\bf Implications and Conclusions.}
IceCube discovered astrophysical neutrinos via an abundance of $\lesssim\,$PeV events.
Even a single highly-energetic $E_\nu \!\gtrsim\! 10$~PeV event is a first direct hint of neutrinos beyond the Glashow resonance, though a deficit of $\sim\,$6~PeV Glashow showers precludes a simple power-law description spanning these regimes.
A tau track event would give insight into the astrophysical neutrino spectrum approaching $E_\nu \!\sim\! 100$~PeV.

{\it Whither Glashow?:}
A ``successful'' model should yield sufficient track rates to account for the event depositing 2.6~PeV, without overproducing multi-PeV showers.
The rates from our nominal $\varphi_i$ models are in plausible ranges to source a track event; however, puzzles remain.

$\varphi_7$: The minimal model such to yield $E_\mu \!\gtrsim\! 5$~PeV muons, though disfavored at $\gtrsim$99\% by Glashow rates unless the normalization is greatly reduced.  This would suppress track rates.

$\varphi_8$: Yields fewer muons than $\varphi_7$, though much fewer Glashow events and a sizable $\tau$-track fraction.  We find via a likelihood calculation that $\varphi_8$ with a slightly decreased normalization is most favored \cite{SuppM}.  A tau track identification would point to such a model.

$\varphi_9$: Though less likely for $\sim$2.6~PeV tracks, shower rates are small.
The upgoing tau spectrum peaks at $E_\tau \!\sim\! 200$~PeV.
We note an ANITA $600 \!\pm\! 400$~PeV shower event could be an upgoing tau decaying above the ice, though at $\sim$$20^\circ$ upgoing is perplexing \cite{Gorham:2016zah}.
While $\varphi_9$ itself is viable, an accompanying $\varphi_7$-like GZK flux \cite{SuppM} disfavors many combinations.

We find that $E_\nu^{-2.6}$ is disfavored at the $\sim$90\% level due to low track rates. 
We also find that Glashow rates (Table~\ref{tab:rates}) disfavor the best fit $E_\nu^{-2.13}$ spectrum (cutoff at 10~PeV; Fig.~\ref{casca}) from IceCube muon studies \cite{Aartsen:2016xlq} at $\gtrsim$99\% \cite{SuppM}.
Intermediate models $E_\nu^{-2.13} \, {\rm exp} [-E_\nu/6.9 \, {\rm PeV}]$ or $E_\nu^{-2.6}$ cutoff at 10 PeV perform no better (in Table~\ref{tab:rates} models ``$E_\nu^{-2.13}$e'' and ``$E_\nu^{-2.6}$c'', respectively; see \cite{SuppM}).
Importantly, examining muons alone cannot account for the Glashow shower deficit, while pure power-law fits miss spectral transitions.

In IceCube-Gen2 \cite{Aartsen:2014njl,Blaufuss:2015dwa} Glashow shower rates can be $\sim$20$\times$ higher.
Many through-going tau tracks in IceCube would instead be contained, resolving more distinctive topologies \cite{Learned:1994wg,Beacom:2003nh,DeYoung:2006fg}.
An extended surface array \cite{Jero:2015awa} allows greater veto coverage for downgoing tracks \cite{Euler:2015dwa}.
Such combinations would discriminate \cite{Kistler2014,Laha2013,Anchordoqui2014,Chen:2014gxa} between intrinsically small trans-Glashow fluxes and exotic scenarios, such as
cooled-muon models yielding neutrino spectra from $\pi^+$ decays with $\varphi_{\nu} \!\gg\! \varphi_{\bar{\nu}}$ and negligible Glashow rates (see \cite{Kistler2014}).

{\it Standard Model and Beyond:}
While we quote event rates for all low-background directions, the $2.6 \!\pm\! 0.3$~PeV track comes from a relatively-large angle below the horizon.  This becomes suspicious if similar tracks are not soon detected from downgoing and shallower angles.
We have seen that the cutoffs in Fig.~\ref{angspec} angular distributions are flattened if Earth opacity is decreased.  This could arise from new physics or if $\sigma_{\rm CC}(E_\nu)$ saturates at $\gtrsim\,$PeV due to small-$x$ QCD effects \cite{Henley:2005ms}.

New-physics effects are also confronted.  
E.g., for Lorentz invariance violating scenarios \cite{Stecker:2014xja} the multi-PeV track significantly extends previous bounds.

\begin{figure}[t!]
\hspace*{-0.4cm}
\includegraphics[width=1.05\columnwidth,clip=true]{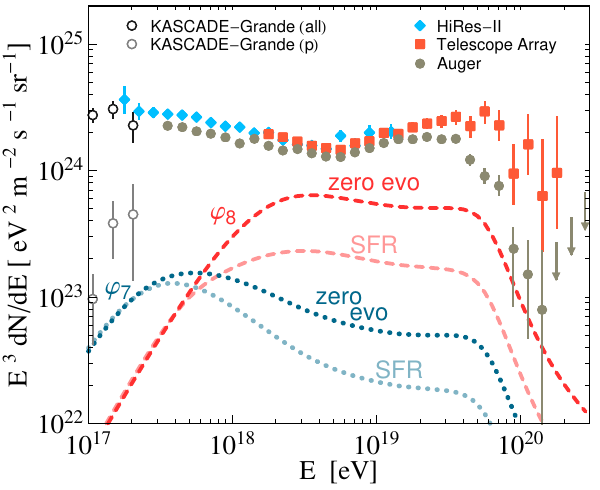}
\caption{Ultrahigh-energy cosmic-ray data \cite{Apel:2013dga,Abbasi:2007sv,Abreu:2011pj,AbuZayyad:2012ru} and proton fluxes associated with neutrino Models $\varphi_7$ ({\it dotted}) and $\varphi_8$ ({\it dashed}) assuming zero ({\it dark}) or star formation rate ({\it light}) source evolution.
\label{UHECR}}
\end{figure}

{\it UHECR Connections:}
For our neutrino emissivities \cite{SuppM}
we assume $\pi^\pm$$\mu^\pm$ decays yield six neutrinos for each neutron of $E_n\!\sim\!20\, E_\nu$ decaying to a proton with $E_p\!\approx\!E_n$ \cite{Kistler2014}.  Taking optically-thin sources, such as BL~Lacs \cite{Padovani:2015mba} motivating $\varphi_7$ and $\varphi_8$, we calculate proton spectra  \cite{Kistler2014}, imposing no cutoff to the high-energy $\beta \!=\! -3$ spectrum.  We do not use $\varphi_9$ (motivated by GZK neutrinos and thus implicitly connected to UHECR).

Fig.~\ref{UHECR} shows the UHECR proton flux from $\varphi_7$ and $\varphi_8$ for zero evolution, as often assumed for BL~Lacs, or cosmic star formation rate  \cite{Hopkins2006,Yuksel2008,Kistler2013b} evolution.  These fall below the data \cite{Apel:2013dga,Abbasi:2007sv,Abreu:2011pj,AbuZayyad:2012ru}, though $\varphi_8$ is close at $\gtrsim\! 10^{18}\,$eV where the composition is light \cite{Abbasi:2004nz,Abbasi:2009nf,Abraham:2010yv}.  Fewer pions per neutron would raise the flux \cite{Kistler2014}, though saturation would leave no room for UHECR mechanisms besides neutron escape from IceCube sources.

{\it Conclusions.}
The $E_{\rm dep} \!\approx\! 2.6$~PeV IceCube track event implies the highest $E_\nu$ interaction to date.
If this track is from a muon, it may indicate a $\gtrsim\,$10~PeV neutrino energy.  Alternatively, we find through-going taus leaving such tracks imply neutrino energy in the $\sim$100~PeV range, giving a glimpse of astrophysical neutrinos from unexpectedly-high energies.

Our calculations show such tracks are unlikely from extending a soft neutrino flux yielding the $\gtrsim$40~TeV IceCube events.
Fluxes like the $\sim$$E_\nu^{-2.1}$ spectrum from analyses of IceCube muons alone imply excessive Glashow shower rates.
We conclude that this combination of low track rates from soft spectra and a deficit of $\sim$6~PeV shower detections favors a new hard astrophysical neutrino flux beyond the Glashow resonance.

The huge separation of parent $\nu_\mu$/$\nu_\tau$ energies producing a through-going track depositing the same energy highlights the importance of developing charged lepton flavor identification for individual tracks.
The models that we considered suggest the IceCube multi-PeV track is the tip of a super-Glashow iceberg and
detectors such as IceCube Gen-2 \cite{Aartsen:2014njl}, ARIANNA \cite{Barwick:2014pca}, and ARA \cite{Allison:2011wk}
can improve prospects of addressing flavor ratios, the birthplaces of UHECR, and more.
\\

\acknowledgments
We thank Markus Ahlers, Basudeb Dasgupta, Tyce DeYoung, Amol Dighe, Alex Friedland, Tom Gaisser, Raj Gandhi, Francis Halzen, Naoko Kurahashi-Neilson, Claudio Kopper, John Learned, Shirley Li, Kohta Murase, Hans Niederhausen, Kenny C.Y.\ Ng, Dave Seckel, Justin Vandenbroucke, Nathan Whitehorn, Donglian Xu, and especially John Beacom and Hasan Yuksel for discussions and INT Program INT-15-2a ``Neutrino Astrophysics and Fundamental Properties'' for hospitality early in this project.
MDK and RL acknowledge support provided by Department of Energy contract DE-AC02-76SF00515 and the KIPAC Kavli Fellowship made possible by The Kavli Foundation.
RL is supported by the German Research Foundation (DFG) under Grants No.~EXC-1098, No.~KO 4820/1-1, and No.~FOR 2239, and from the European Research Council (ERC) under the European UnionÕs Horizon 2020 research and innovation program (Grant No.~637506, ``$\nu$Directions'') awarded to Joachim Kopp.





\clearpage
\newpage
\maketitle
\onecolumngrid
\begin{center}
\textbf{\large Multi-PeV Signals from a New Astrophysical Neutrino Flux Beyond the Glashow Resonance} \\ 
\vspace{0.05in}
{ \it \large Supplementary Material}\\ 
\vspace{0.05in}
{ Matthew D. Kistler and Ranjan Laha}
\end{center}
\onecolumngrid
\setcounter{equation}{0}
\setcounter{figure}{0}
\setcounter{table}{0}
\setcounter{section}{0}
\setcounter{page}{1}
\makeatletter
\renewcommand{\theequation}{S\arabic{equation}}
\renewcommand{\thefigure}{S\arabic{figure}}
\renewcommand{\thetable}{S\arabic{table}}

\begin{figure*}[h]
\vspace*{-0.3cm}
\includegraphics[width=0.9\columnwidth,clip=true]{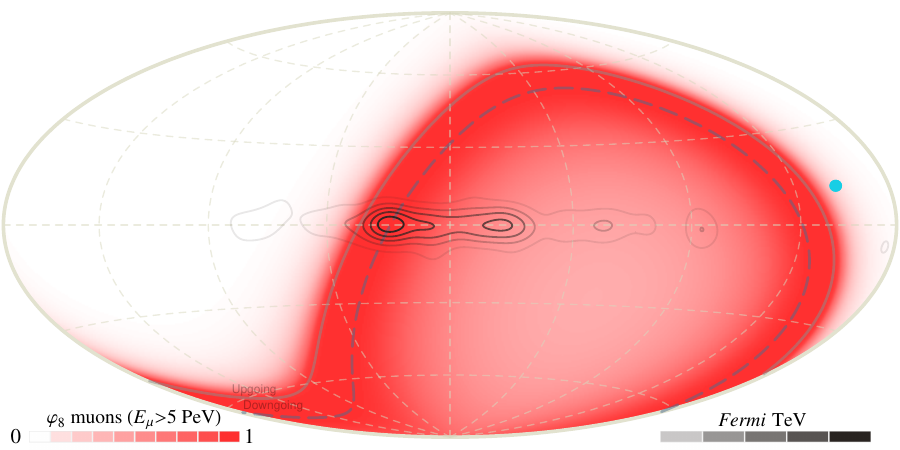}
\vspace*{-0.cm}
\caption{Sky density of $E_\mu \!>\! 5$~PeV muons from model $\varphi_8$ (in Galactic coordinates; {\it shaded}).  The horizon ({\it solid}) demarcates upgoing and downgoing directions, with the rough $10^\circ$ downgoing boundary for atmospheric muons ({\it dashed}).
The $E_{\rm dep} \!\approx\! 2.6$~PeV IceCube track event ({\it blue dot}) and contours of $E_\gamma \!>\! 1$~TeV {\it Fermi} emission smoothed by $5^\circ$ from \cite{Kistler2015d} are shown for reference.
\label{icetev}}
\end{figure*}


{\bf Multi-PeV Origins:}
Fig.~\ref{icetev} shows the muon sky density arising from our $\varphi_8$ flux model, with a dashed curve $10^\circ$ above the horizon.
We have assumed isotropic neutrino fluxes here and elsewhere, although if the multi-PeV IceCube track arose from emission within the Milky Way or a particularly bright source conclusions could be rather different.

Could this event actually be Galactic?  $E_\nu \!\gtrsim\! 10$~PeV implies a proton energy $E_p \!\gtrsim\! 10^{17}$~eV, well beyond the cosmic-ray proton knee (for nuclei of mass number $A$, we would need to consider $E/A$).
If such neutrino emission resembles that of TeV gamma rays in {\it Fermi} (\cite{Kistler2015d}; see Fig.~\ref{icetev}), we would expect a much higher rate nearer the Galactic Center (GC).
Such a flux gradient outwards from the GC also increases the expected downgoing/upgoing ratio due to the location of IceCube.
That being said, while a location $\sim\!12^\circ$ from the Galactic plane does not indicate Galactic emission,
it is somewhat unlikely if projecting a $\sim\!\pm 10^\circ$ band around the IceCube horizon.

BL~Lac origins have been discussed for each of the three $E_{\rm dep} \!\approx\!  1\!-\!2$~PeV shower events (e.g., \cite{Krauss:2014tna,Padovani:2014bha,Adrian-Martinez:2015jlj,Kadler:2016ygj}), though the angular resolution of such events is limited to $\gtrsim\! 10^\circ$.  However, the 2.6~PeV track is localized to $1^\circ$ at 99\% uncertainty \cite{Schoenen2015,Aartsen:2016xlq}.

Around the best-fit IceCube track position we do not identify any notable object within $1^\circ$.  
At $\sim\!2^\circ$ is PMN~J0717+0941, a relatively-nearby ($\sim\,$123~Mpc) radio galaxy \cite{vanVelzen:2012fn}.  At $\sim\!3^\circ$ is the nearest {\it Fermi} BL Lac, 4C~14.23 \cite{Acero2015}.
No gamma-ray source was reported by HAWC \cite{Taboada2015}.
This event could be from a faint source, though there is no obvious indication of a prominent super-Glashow neutrino source that would violate an assumption of a diffuse, perhaps cosmogenic, flux.

{\bf Multi-PeV Fluxes:}
We formed neutrino spectra $\varphi_i$ from smoothly-broken power laws for simplicity, not attempting to match details of specific models.
For instance, $\alpha \!=\! -1$ in Eq.~(\ref{specfit}) is somewhat softer than typical GZK fluxes from interactions with the CMB (see, e.g., Fig.~4 of \cite{Yuksel2007}), although including the EBL leads to further softening.  We also chose $\beta \!=\! -3$ for a decline, though we could have used exponential cutoffs, the spectral peaks being most relevant for event rates.  For convenience, the normalizations in Eq.~(\ref{specfit}) are $f_7 = 3 \times 10^{-22}$, $f_8 = 3 \times 10^{-24}$, and $f_9 = 3 \times 10^{-26}$, all in terms of GeV$^{-1}\,$cm$^{-2}\,$s$^{-1}\,$sr$^{-1}\,$.

Cosmic neutrino emissivities for each $\varphi_i$ can be obtained from a suitable source $dN_\nu/dE_\nu$ as
\es{f2}{
\varphi_\nu(E_\nu) = \frac{c}{4 \pi } \int_0^{z_{\rm max}} \frac{dN_{\nu}}{dE_{\nu}^\prime}  \frac{dE_{\nu}^\prime}{dE_{\nu}}\, \frac{\mathcal{W}(z)}{dz/dt} \,dz \,,}
where ${dz}/{dt} \!=\! H_0\, (1 \!+\! z) [\Omega_m (1 \!+\! z)^3 \!+\! \Omega_\Lambda ]^{1/2}$, ($\Omega_m \!=\! 0.3$, $\Omega_{\Lambda} \!=\!0.7$, and ${H}_{0} \!=\! 70\,$km/s/Mpc), and $dE_\nu^\prime/dE_\nu \!=\! (1+z)$, accounting for source evolution with redshift, $\mathcal{W}(z)$, with appropriate adjustments of the spectral parameters (see \cite{Kistler2015c}).

\begin{figure*}[t!]
\vspace*{-0.2cm}
\hspace*{-0.4cm}
\includegraphics[width=0.5 \columnwidth,clip=true]{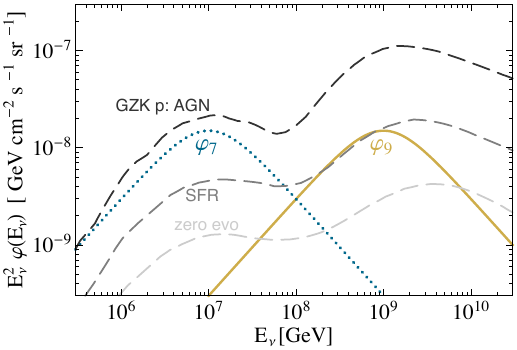}
\includegraphics[width=0.5 \columnwidth,clip=true]{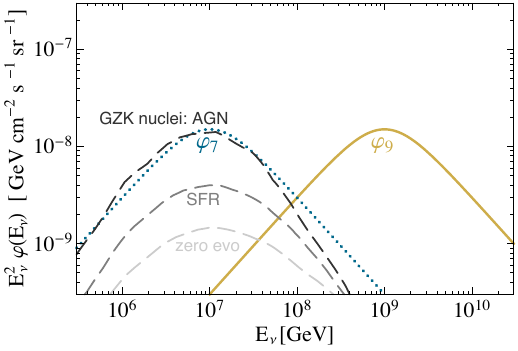}
\vspace*{-0.4cm}
\caption{
{\it Left:} Cosmogenic neutrino models ({\it dashed lines}) for either AGN, SFR, or zero cosmic evolution assuming a proton dominated UHECR composition up to $\gtrsim\! 10^{20}$~eV \cite{Aloisio:2015ega}.  The peak at $\sim\!10^7$~GeV is due to interactions with the EBL, while the $\gtrsim\!10^9$~GeV peak is from the CMB.  These are compared to our fiducial models $\varphi_7$ and $\varphi_9$ from Fig.~\ref{casca}.  For example, $\varphi_{\rm SFR} \approx 0.3\, \varphi_7 +  \varphi_9$.
{\it Right:} The same for cosmogenic neutrinos assuming a mixed nuclear composition with the same evolutions \cite{Aloisio:2015ega}.  These have a shape strongly resembling our $\varphi_7$.}
\label{compspec}
\end{figure*}

As mentioned in the main text, Model $\varphi_9$ approximates the cosmogenic spectrum from $p \gamma$ interactions on the CMB, with the shape of $\varphi_7$ roughly that of lower-energy proton interactions with the extragalactic background light (EBL), so that $\varphi_{\rm GZK} \approx C_7 \varphi_7 + C_9  \varphi_9$.
The particular combination in Fig.~\ref{casca} does not correspond to any physical model, though rescaling can approximate various scenarios.
Fig.~\ref{compspec} shows models from \cite{Aloisio:2015ega} assuming either proton-dominated ({\it Left}) or mixed nuclear ({\it Right}) UHECR compositions up to $E_p \!\gtrsim\! 10^{20}$~eV.
We see that, e.g., $C_7 \sim 0.3$ and $C_9 \sim 1$ approximates proton-dominated SFR source evolution.  Though the $\varphi_9$ shape somewhat underestimates fluxes at $E_\nu \!\gtrsim\! 10^{9}$~GeV for these particular models from Ref.~\cite{Aloisio:2015ega}, these energies are not of great relevance for our present study.
Using $C_7 \sim 1$ and $C_9 \sim 0$ incidentally very-well describes a mixed nuclear composition with AGN evolution.  Other combinations can be made as desired to examine neutrino event rates using the values in the main text.

\begin{figure*}[b]
\vspace*{-0.3cm}
\includegraphics[width=0.6\columnwidth,clip=true]{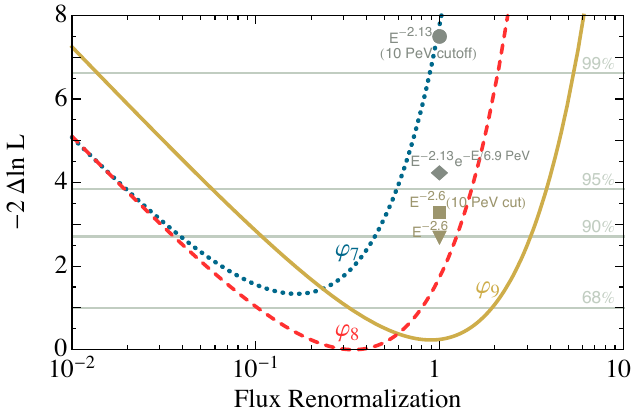}
\vspace*{-0.cm}
\caption{Likelihood analysis based on Eq.~(\ref{f3}) with varied the flux normalizations for models $\varphi_{7}$, $\varphi_{8}$, and $\varphi_{9}$.
We find that intermediate models, such as $\varphi_{7.5}$ or $\varphi_{8.5}$ (not shown), reside between these curves.
We also show values obtained from several power-law models with fixed normalization.
\label{likfig}}
\end{figure*}

{\bf Statistics:}
We formed neutrino spectra $\varphi_i$ from smoothly-broken power laws for simplicity, not attempting to match details of specific models.
This allows for freedom to examine whether fluxes in various energy ranges are most likely to account for high-energy neutrino data.

The Poisson likelihood for each $\varphi_i$ can be obtained via standard techniques (e.g., \cite{Chen:2014gxa}) as
\es{f3}{
\mathcal{L}_i = \prod_{j}  \frac{e^{-\lambda_j} \lambda_j^{n_j}}{n_j !}  \,, }
%
where $j$ corresponds to energy bins containing measured counts $n_j$ and predicted event rates based on exposure and methods discussed in the main text.

Exact values will of course vary depending upon choices of binning and exposure.
The IceCube High-Energy Starting Event (HESE) analysis focussed on finding lower-energy events down to $\lesssim\,$100~TeV, with the effective volume for $\nu_e$ showers already saturating at $\sim\,$0.4~km$^3$ by $\sim\,$200~TeV \cite{Aartsen2013b}.  Meanwhile, their Extremely-High Energy (EHE) search is principally looking for cosmogenic neutrinos with $\gtrsim\,$EeV energies, with cuts used by the latest EHE analysis leading to the exclusion of even the famed $1-2$~PeV showers (``Bert'', ``Ernie'', and ``Big Bird'') from the EHE event sample \cite{Aartsen:2016ngq}.  Unfortunately, the Glashow energy range does not appear to have yet been optimized in an existing search.  It does appear very likely, though, that a shower with $\gtrsim\,$6~PeV energy deposition would be detected within the $\sim\,$1~km$^3$ instrumented IceCube volume in the last $\sim\,$7~yr of operation.  We use nominal values of 5~km$^2\,$yr for tracks and 5~km$^3\,$yr for showers to estimate the present sensitivity, though it is straightforward to consider a range if effective areas and volumes can be simply rescaled by a common factor in forming exposures.

We consider here rates of all events with $E_\mu > 3$~PeV, $E_{\rm em} > 3$~PeV, and $E_\tau > 50$~PeV.  We use bins covering $3 < E_\mu < 30$~PeV, $E_\mu > 30$~PeV, $50 < E_\tau < 200$~PeV, $E_\tau > 200$~PeV, and $E_{\rm em} > 3$~PeV.  These bins are motivated by the observed track data and the existing degeneracy between muon and tau tracks in this range, for which we thus sum the muon and tau track rates.  Fig.~\ref{likfig} shows the likelihoods obtained from varying the nominal flux normalization of each model considered.  Most notably, $\varphi_{8}$ with a slightly lower normalization than our nominal value is most preferred, while our nominal $\varphi_{7}$ model is strongly disfavored.   Even decreased $\varphi_{7}$ normalizations do not fare well because of the predicted Glashow shower rates.  If considered alone, $\varphi_{9}$ is quite consistent, though GZK scenarios with a significant $\varphi_{7}$-like component are degraded accordingly.  An intermediate ``$\varphi_{8.5}$'' model would also fare very well on its own.

Using the same bins and exposure, we calculate values for several models based on power-law spectra with various cutoff assumptions.
We note that extending down to $E_\mu = 3$~PeV, rather than $E_\mu = 5$~PeV, actually advantages these models due to their softer spectra, yet still does not compensate for their low rate of tracks and/or high rate of showers.  The power law models are taken to have a fixed normalization and we show that they are a poorer fit to the data in the energy region that we consider.  Further data will clarify the neutrino flux at these high energies.   Upcoming radio array experiments like ARIANNA will also have the sensitivity to probe this interesting region\,\cite{Barwick:2014pca}.

{\it Note added:}  IceCube collaboration presented new analysis in ICRC 2017\,\cite{Aartsen:2017mau}.  They did not detect any new through going track event with deposited energy $\gtrsim$ PeV and the HESE search did not detect any new event with deposited energy $\gtrsim\,$200 TeV.  These new results do not affect our conclusions.

\end{document}